\documentclass[3p,times,procedia,preprintnumbers,amsmath,amssymb,nofootinbib,epsfig,bmamsfonts,yfonts,superscriptaddress]{elsarticle}
\usepackage{color}

\usepackage{amssymb}
\usepackage{amsmath}

\makeatletter
\newcommand{\xRightarrow}[2][]{\ext@arrow 0359\Rightarrowfill@{#1}{#2}}
\makeatother

\newcommand{\Trento}{${\rm T_RENTo}$\,}

\usepackage[figuresright]{rotating}

\begin{document}

\begin{frontmatter}

\title{Hydrodynamization in systems with detailed transverse profiles}
%

\author{Aleksi Kurkela$^{1,2}$, Seyed Farid Taghavi$^{3}$, Urs Achim Wiedemann$^{1}$ and Bin Wu$^{1}$}

\address{$^1$ Theoretical Physics Department, CERN, CH-1211 Gen\`eve 23, Switzerland\\
$^2$ Faculty of Science and Technology, University of Stavanger, 4036 Stavanger, Norway\\
$^3$ Physik Department E62, Technische Universit\"{a}t M\"{u}nchen, James Franck Str.~1, 85748 Garching, Germany}

\begin{abstract}
The observation of fluid-like behavior in nucleus-nucleus (AA), proton-nucleus (pA) and high-multiplicity proton-proton (pp) collisions
motivates systematic studies of how different measurements approach 
their fluid-dynamic limit. We have developed numerical methods to solve the ultra-relativistic Boltzmann equation for systems
of arbitrary size and transverse geometry. Here, we apply these techniques for the first time to the study of azimuthal flow coefficients $v_n$ including non-linear mode-mode coupling and to an initial condition with realistic event-by-event fluctuations.
We show how both linear and non-linear response coefficients extracted from $v_n$ develop as a function of opacity
from free streaming to perfect fluidity. We note in particular that away from the fluid-dynamic limit, the signal strength of linear and non-linear response coefficients does not reduce uniformly, but that their hierarchy and relative size shows characteristic differences.
\end{abstract}


\end{frontmatter}
\vspace{0.5cm}

{\bf Introduction.}
\emph{Hydrodynamization} denotes the transition to hydrodynamics of systems that carry
 fluid- and non-fluid-dynamic degrees of freedom and that therefore do not need to behave fluid dynamically at all 
times and under all conditions. The observation of strong signs of collectivity in ultra-relativistic nucleus-nucleus (AA), 
proton-nucleus (pA) and proton-proton (pp) collisions \cite{Abelev:2014mda,Sirunyan:2017uyl,Aaboud:2017acw} 
has motivated in recent years many studies of \emph{hydrodynamization} 
in strongly- and weakly-coupled models of quark-gluon plasma
~\cite{Chesler:2008hg,Heller:2011ju,Kurkela:2015qoa,Keegan:2015avk,Berges:2020fwq,Heller:2015dha,Heller:2016rtz,Heller:2013fn,Romatschke:2017vte,Kurkela:2019set,Strickland:2017kux,Blaizot:2017ucy,Almaalol:2020rnu,Spalinski:2018mqg,Behtash:2019txb,Strickland:2019jut,Chesler:2015wra,Denicol:2018pak,Bantilan:2020pay,Attems:2017zam,Attems:2018gou,Kurkela:2018qeb,Kurkela:2019kip,Heller:2018qvh,Basar:2015ava,Spalinski:2017mel,Behtash:2017wqg,Brewer:2019oha}. Their ultimate aim is to provide a rigorous underpinning of the fluid-dynamic interpretation of collective flow in AA, pA and pp collisions, and to delineate the limitations of any such interpretation.  

 Most studies of hydrodynamization profit from simplified set-ups that do not reflect all phenomenological 
 complications but that exhibit general features in great clarity. In particular, most studies of hydrodynamization to date 
 assume exact Bjorken boost invariance, employ conformally symmetric collective dynamics and focus on 
 dimensionally reduced $1+1$D systems~\cite{Chesler:2008hg,Heller:2011ju,Kurkela:2015qoa,Keegan:2015avk,Heller:2015dha,Heller:2016rtz,Heller:2013fn,Romatschke:2017vte,Strickland:2017kux,Blaizot:2017ucy,Almaalol:2020rnu,Spalinski:2018mqg,Strickland:2019jut}
(for studies extending this framework, see~\cite{Kurkela:2019set,Behtash:2019txb,Chesler:2015wra,Denicol:2018pak,Bantilan:2020pay,Attems:2017zam,Attems:2018gou,Kurkela:2018qeb,Kurkela:2019kip,Keegan:2016cpi}).
 Within this setting, one has reached in recent years
 a thorough understanding of the off-equilibrium evolution of simple observables in various models. For instance, 
 the asymmetry $p_T/p_L$ between longitudinal and transverse pressure and the higher longitudinal 
 momentum moments of the stress-energy tensor are known to approach rapidly their universal attractor solution in kinetic theory~\cite{Strickland:2018ayk, Kurkela:2019kip,Almaalol:2020rnu}. 
 The mathematical structures behind this behaviour continue to be studied in the context of resurgence~\cite{Heller:2015dha,Heller:2016rtz,Heller:2018qvh,Basar:2015ava}.

The lessons learnt from these $1+1$D systems are expected to carry over to the phenomenological reality in $3+1$D. 
For instance, the early-time dynamics of $p_T/p_L$ in boost-invariant $3+1$D systems is known to be governed
locally in the transverse plane by an effective $1+1$D evolution, and the $1+1$D universal attractor for $p_T/p_L$ is
therefore of relevance for the $3+1$D dynamics. However, very few observables of phenomenological
relevance can be studied in $1+1$D systems, and some important questions have therefore received little attention
so far in the debate of hydrodynamization.  One of them is whether all bulk observables hydrodynamize 
under conditions comparable to those under which $p_T/p_L$ hydrodynamizes, or whether some classes of observables 
require systems of longer lifetime, larger spatial extent and/or higher density to approach the values they attain under 
conditions of almost perfect fluidity. Of particular interest in this context are the conditions for hydrodynamization of 
the azimuthal momentum anisotropies $v_n$ of soft multi-particle production, as these are amongst the most abundant
and most precisely measured signatures of collective behavior in AA, pA and pp collisions. Here, we analyze their
hydrodynamization in a boost invariant conformally symmetric $3+1$D kinetic transport theory, whose $1+1$D variants 
have been used repeatedly in studies of hydrodynamization.


Up until this point, only the linear response coefficients have been studied in full kinetic theory because of the 
technical challenges related to solving Boltzmann equations for a distribution functions in complex geometries~\cite{Kurkela:2018ygx,Kurkela:2018qeb,Kurkela:2019kip}, though some results exist for perturbative solutions
around free-streaming~\cite{Borghini:2018xum,Kurkela:2018ygx}. We have developed numerical
techniques to solve such systems and we present here the first non-linear response coefficients, and we present 
the first solution to the Boltzmann equation for an initial condition with realistic event-by-event fluctuations.

{\bf Kinetic Theory}.
We consider massless, boost-invariant kinetic theory in the isotropization-time approximation, and we restrict the discussion
to the first momentum moments $F(\vec x_\perp,\Omega,\tau) = \int\frac{4 \pi p^2 d p}{(2\pi)^3} p f$ 
of the distribution function $f$. Here, $p$ is the modulus of the three-momentum, the velocity is
$v_\mu \equiv p_\mu/p$ with $p_\mu\, p^\mu = 0$ and $v^0 = 1$, and $\Omega$ denotes the angular phase space of $v_\mu$.
$F$ defines the energy momentum tensor 
$T^{\mu\nu} = \int d\Omega\, v^\mu v^\nu F$, as well as arbitrary higher $v_\mu$-moments that lie beyond hydrodynamics. 
It satisfies the equations of motion~\cite{Kurkela:2019kip}
\begin{equation}
\partial_\tau F + \vec{v}_\perp \cdot \partial_{\vec{x}_\perp} F - \frac{v_z}{\tau}(1-v_z^2) \partial_{v_z} F + \frac{4 v_z^2}{\tau}F 
= -C[F] = -\gamma \varepsilon^{1/4}(x) [-v_\mu u^{\mu} ] (F - F_{\rm iso})\, ,
\label{eq1}
\end{equation}
where $\varepsilon$ is the local energy density. Fluid-like and particle-like excitations are known to coexist in this kinetic 
transport and their properties can be calculated analytically. In particular,
the coupling $\gamma$ is related to the specific shear viscosity $\frac{\eta}{sT} = \textstyle\frac{1}{5 \gamma\varepsilon^{1/4}}$, and
$F$ relaxes locally on a time scale $\tau_R =\tfrac{1}{\gamma\varepsilon^{1/4}}$ to the isotropic distribution 
$F_{iso}(\tau, \vec x_\perp; \Omega)   = \frac{\varepsilon(\tau, \vec x_\perp)}{(-u_\mu v^{\mu})^4}$ 
whose functional form is fixed by symmetries and by the Landau matching  condition, 
$u^\mu T_{\mu}^{\, \, \,\nu}  = - \varepsilon u^\nu$.

As the dynamics \eqref{eq1} is scaleless, dimensionful characteristics of
the collision system can enter only via the initial conditions, and they can affect results only in dimensionless combinations.
For a system of transverse r.m.s.~size $R$ and energy density $\varepsilon_0$ at initial time $\tau_0$, it follows that the opacity $\hat\gamma=\gamma R^{3/4} (\varepsilon_0\tau_0)^{1/4}$ is the unique model parameter. Eq.~\eqref{eq1} interpolates between
 free-streaming in the limit of vanishing opacity $\hat\gamma\to 0$ and ideal fluid dynamics in the limit  $\hat\gamma \to \infty$. 
 
We initialize \eqref{eq1} with two different classes of initial conditions. We first study
linear and non-linear response coefficients  based on the simple Gaussian ansatz
\begin{equation}
	F(\tau_0, \vec x_\perp; \phi, v_z) = 2 \varepsilon_0\, \delta(v_z)\, \exp\left[ -\frac{r^2}{R^2} \right]\, 
	\left(  1 + \sum_n \delta_n \left(\frac{r}{R}\right)^n \cos\left(n\theta - n\psi_n\right) \exp\left[ -\frac{r^2}{2R^2} \right] \right)\, .
	\label{eq2}
\end{equation}   
The exponential $\exp\left[-\textstyle\frac{r^2}{2R^2}\right]$  multiplying the $\cos$-term ensures that the distribution 
stays positive everywhere for sufficiently small $\delta_n$'s. 
The initial spatial azimuthal asymmetries are proportional to the real factors $\delta_n$, and they are oriented along
the azimuthal directions  $\psi_n$. Alternatively, we initialize \eqref{eq1} also with the ``realistic'' initial conditions arising from the
\Trento model by replacing the radial profile with that arising from the initial state model. 

For both classes of initial conditions, we quantify azimuthal anisotropies in terms of the complex-valued spatial eccentricities
for $n>1$, 
\begin{equation}
\epsilon_{n} \equiv -\frac{
\int d\theta \, r\, dr\, r^n \exp\left[ i n\theta\right]\, F(\tau_0, \vec x_\perp;\Omega)}{ \int d\theta \, r\, dr\, r^n  F(\tau_0, \vec x_\perp;\Omega)}
\equiv \vert \epsilon_{n} \vert\, e^{i\, n\, \psi_n}\, .
\label{eq3}
\end{equation}
%
Evolving with eq.~\eqref{eq1} the initial conditions \eqref{eq2}, we obtain the evolution of the energy-momentum tensor 
$T^{\mu\nu}$ and the transverse energy flow $d E_\perp$  at late times 
\begin{equation}
 \frac{d E_\perp}{d \eta_s d \phi} \!\equiv\!\! \int\!\! dp^2_\perp   \frac{p_\perp\, d N}{dp^2_\perp d \eta_s d \phi} \!
=\! \frac{d E_\perp}{2\pi d \eta_s}\! \left(\!1\! +\! 2\sum_{n=1}^\infty v_n\cos \left(n\, \phi - n\, \phi_n \right) \right) \, .
\label{eq4}
\end{equation}
This determines the energy flow coefficients 
$V_n = v_n\, e^{i\, n\, \phi_n}$, where $\phi_n$ is the azimuthal orientation of the energy flow. 
In contrast to flow coefficients extracted from particle distributions $dN$,
our study focusses on energy-flow coefficients which are not affected by hadronization since hadronization conserves 
energy and momentum.

The viscous fluid-dynamic limit of eq.~\eqref{eq1} is restricted to the evolution of seven fluid-dynamic fields which may 
be identified with those seven components of $T^{\mu\nu}(\tau, \vec x_\perp)= \int d\Omega\, v^\mu v^\nu F$ that 
do not vanish under boost-invariance. We are interested in the apparently simple kinetic theory \eqref{eq1} 
for $F(\tau, \vec x_\perp; \phi, v_z)$ away from the fluid dynamics limit since it provides an explicit realization of 
fluid fields coupled to a tower of arbitrarily many non-fluid-dynamic excitations (that may be 
parametrized by the higher $v^\mu$-moments of $F$). However, going beyond the fluid-dynamic
limit has a price: $F$ depends on two additional dimensions $\phi$ and $v_z$ in 
momentum space. Discretizing $\phi$ in twenty points and discretizing the 
$v_z$-dependence in 50 points implies a 1000-fold increase of the numerical complexity compared to viscous fluid dynamics. 
The numerical method for solving this evolution equation~\eqref{eq1} has been described in ~\cite{Kurkela:2019kip}, but
there it was applied only to the linear response  of flow coefficients for infinitesimally small $\epsilon_n$ when the coupling 
between different harmonics can be neglected and the numerics simplifies. Here, we overcome this remaining limitations and we study the kinetic theory for arbitrary eccentricities, arbitrary opacities, and arbitrary coupled
non-linear responses.

\begin{figure}[t]
 \includegraphics[width=0.5 \textwidth]{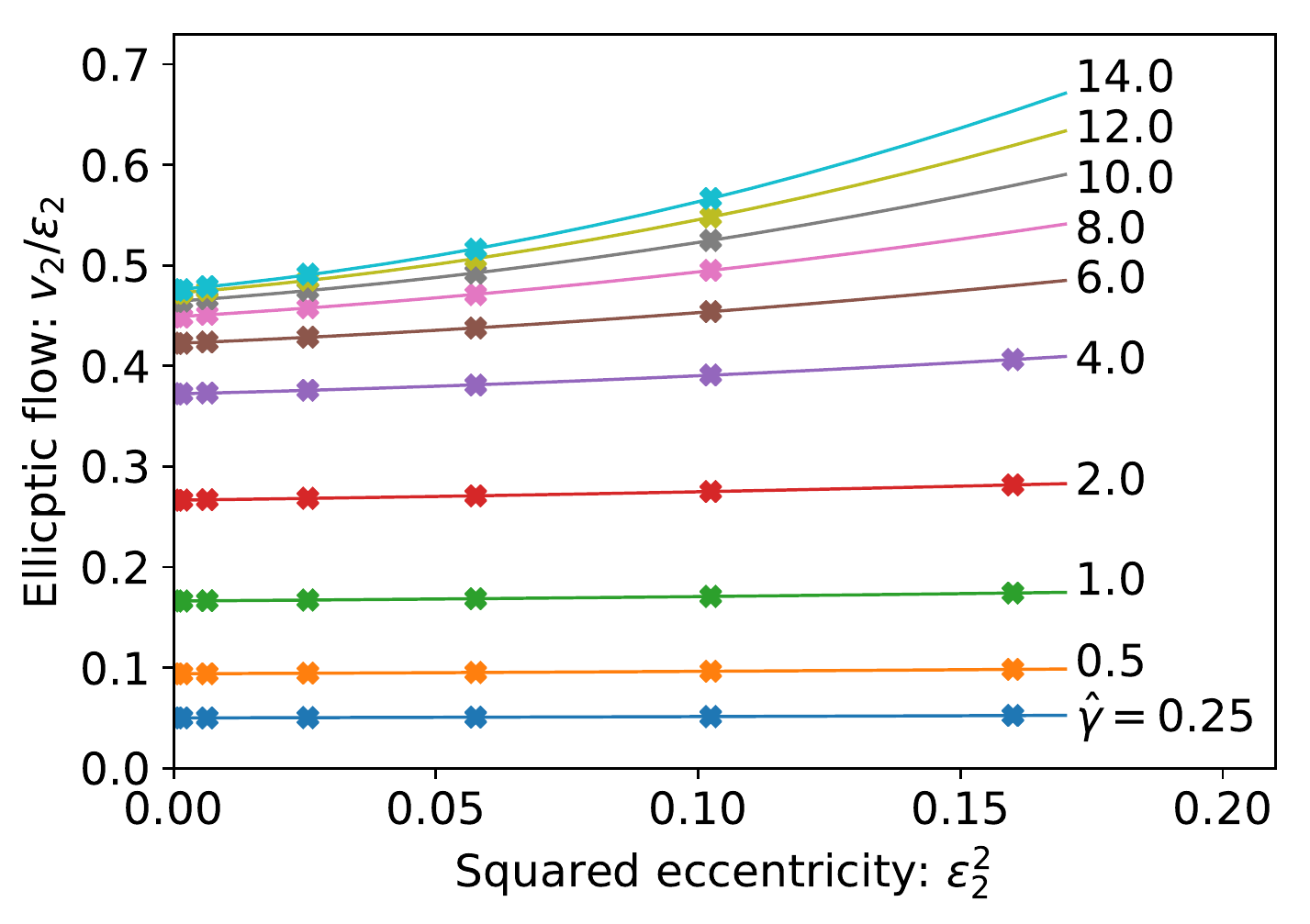}
  \includegraphics[width=0.5 \textwidth]{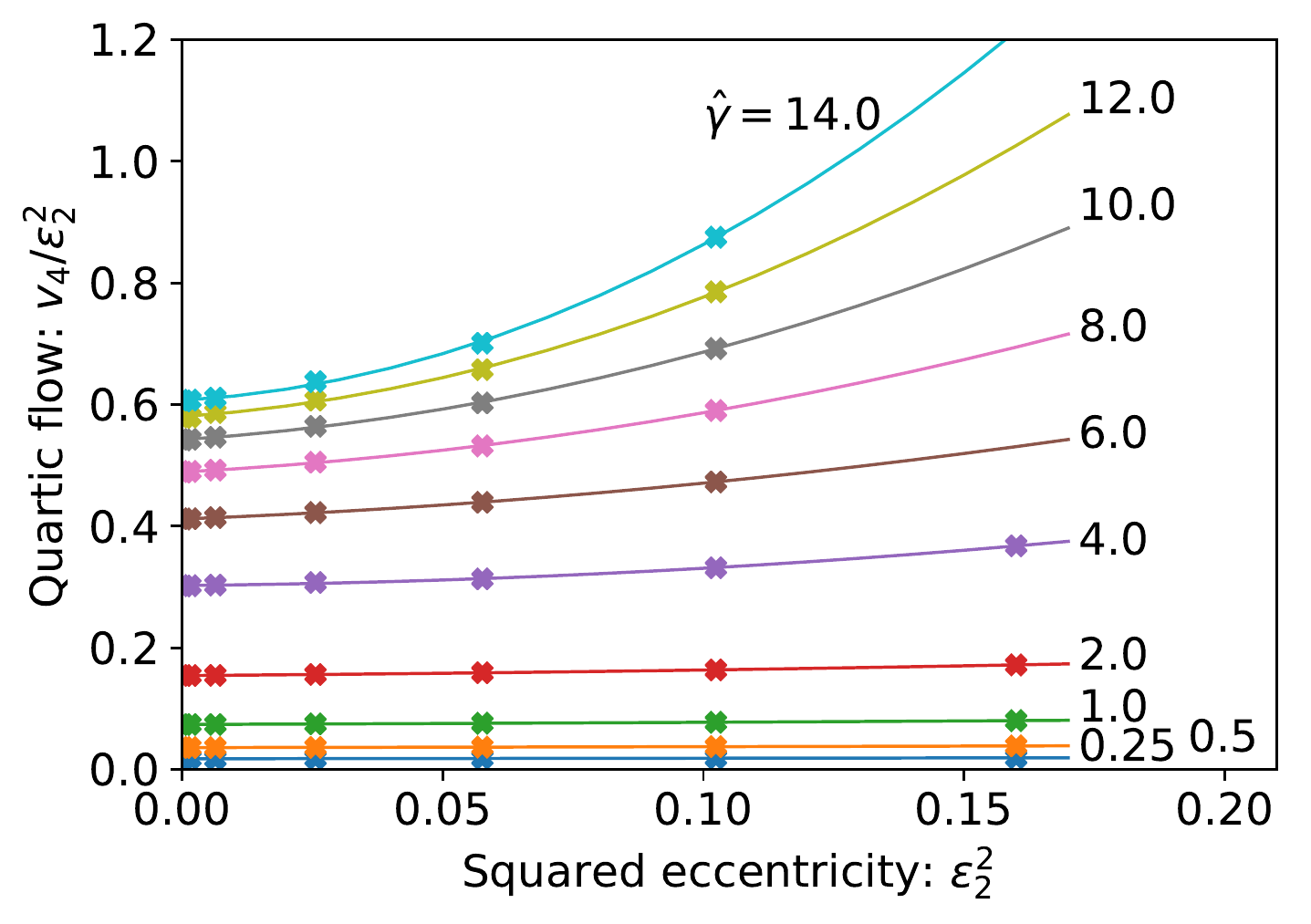}
  \caption{Elliptic $v_2$ (left panel) and quartic $v_4$ (right panel) energy flow coefficients, divided by the leading power $\epsilon_2$  and $\epsilon_2^2$ of the perturbative expansion \eqref{eps2response}, respectively. Results of the kinetic theory \eqref{eq1}, 
  with initial conditions \eqref{eq2} seeded by a single non-vanishing elliptic eccentricity are displayed for different values of
  the opacity $\hat\gamma$ and as a function of the squared eccentricity $ \left( \epsilon_2  \epsilon_2^*\right)$. }
   \label{fig1}
\end{figure}

{\bf Results for mode-by-mode kinetic theory}. 
The coefficients $v_n$ are known to arise from the dynamical response to spatial eccentricities $\epsilon_n$ in the inital 
nuclear overlap. 
The numerically largest responses are linear ($v_n \propto w_{n;n} \epsilon_n$)~\cite{ALICE:2011ab}, but sizeable quadratic  
($\propto  w_{n;n_1,n_2}  \epsilon_{n_1} \epsilon_{n_2}$) and cubic corrections have been quantified~\cite{Acharya:2017zfg,Teaney:2012ke} 
and these can dominate higher harmonics ($n\geq 4$). For linear responses to spatial eccentricities, there is an intrinsic ambiguity between the 
initial geometry that specifies the values $\epsilon_n$, and the collective dynamics that builds up $v_n$ from these 
$\epsilon_n$. Non-linear response coefficients are of particular interest, since they help to disentangle this ambiguity.

\begin{figure}[t]
 \includegraphics[width=0.5 \textwidth]{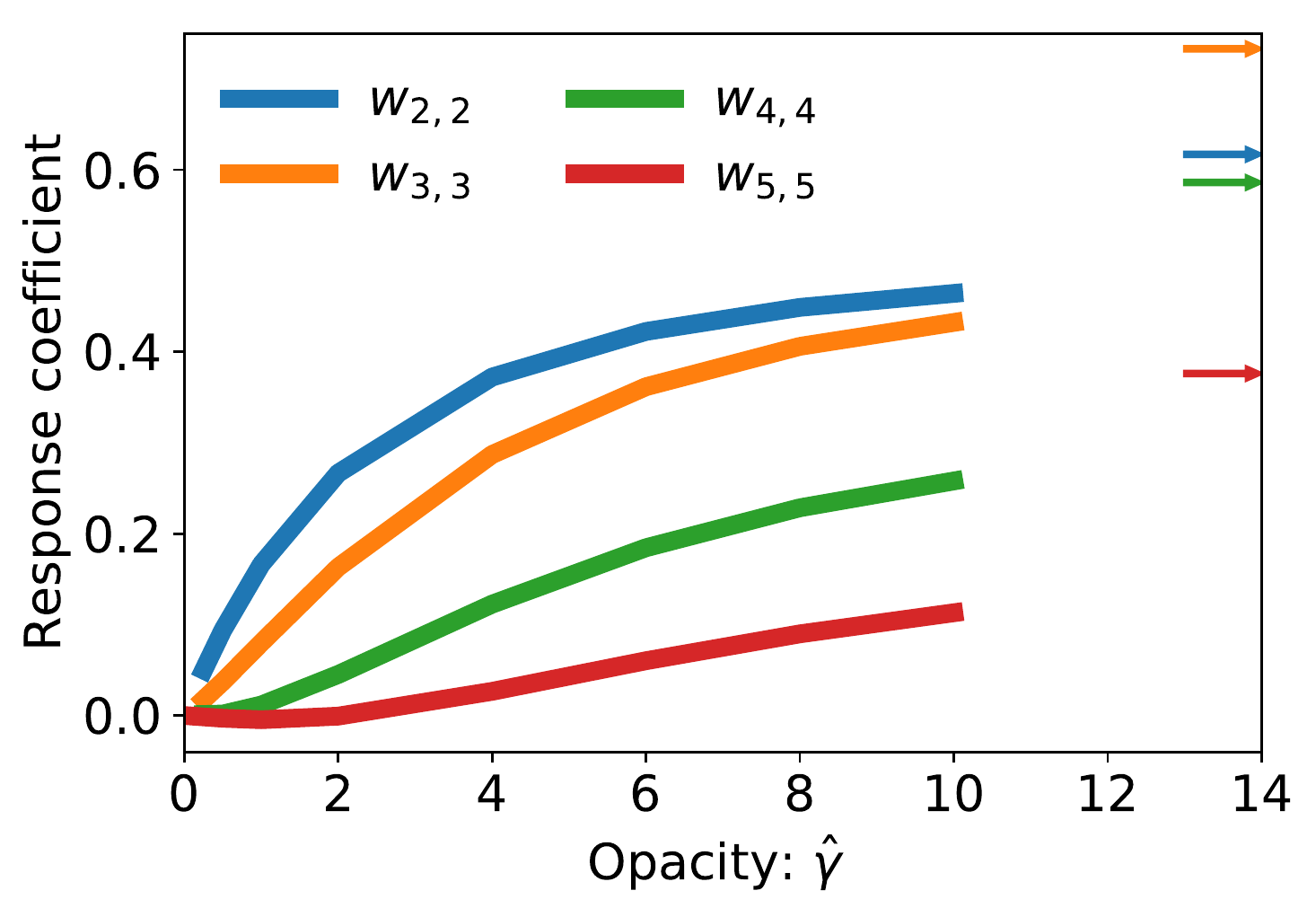}
  \includegraphics[width=0.5 \textwidth]{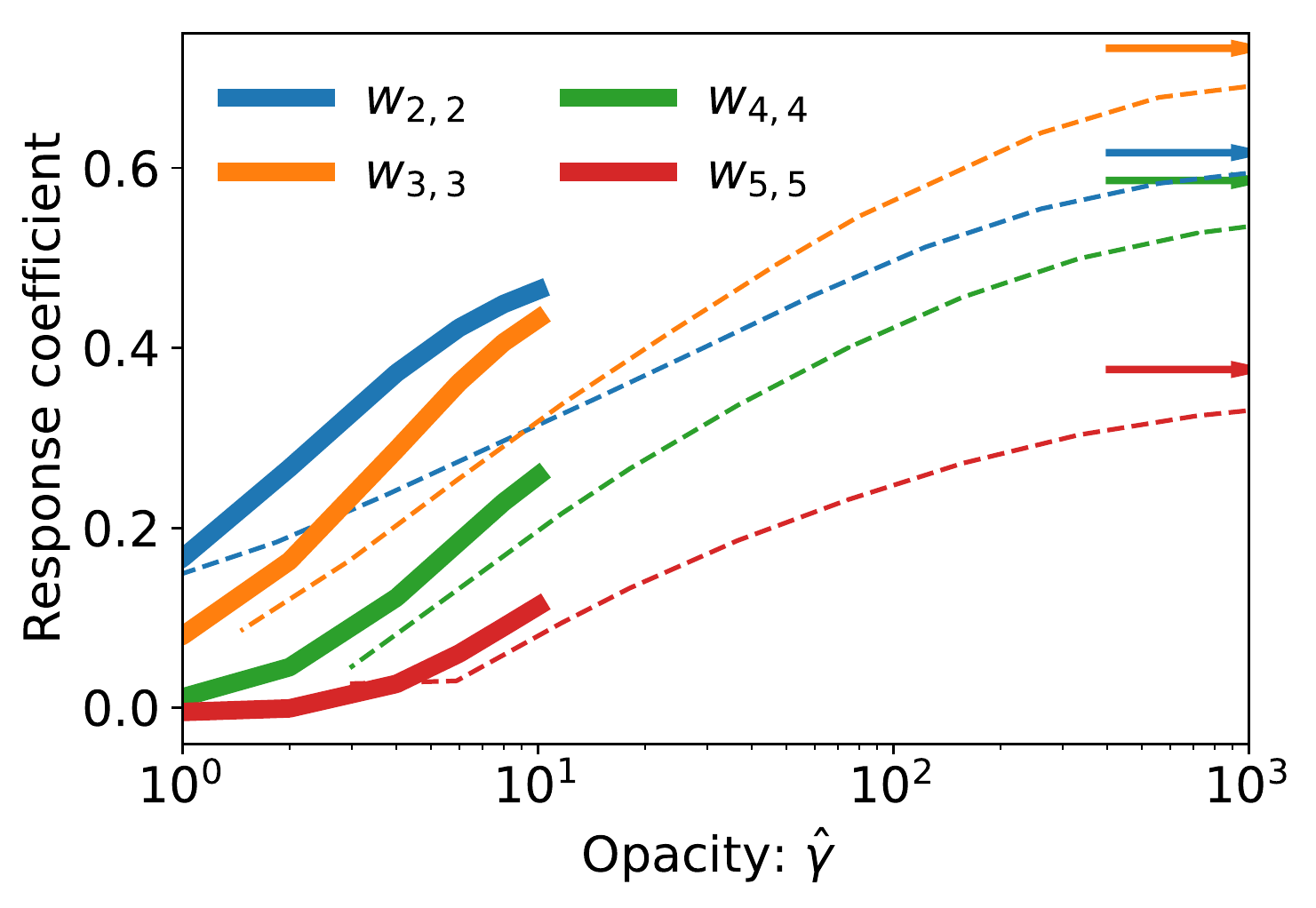}
  \caption{Left panel:  the linear response coefficients 
  $w_{n,n}=\lim_{\epsilon_n \to 0}\textstyle\frac{v_n}{\epsilon_n}$ calculated for the kinetic theory \eqref{eq1}, \eqref{eq2}
  as a function of opacity $\hat\gamma$ (thick lines). Arrows at the right indicate values in the ideal-fluid limit corresponding
  to $\hat\gamma \to \infty$. Right panel: 
  Same as in left panel but in semi-logarithmic presentation and overlaid with results from viscous fluid dynamics
 (thin dashed lines).  }
  \label{fig2}
\end{figure}

As a first example, we consider initial conditions \eqref{eq2} in which a 
single mode $\delta_2$ is excited  ($\delta_n = 0$ for $n\not= 2$). In the course of
the evolution, the non-linear mode-mode coupling of this initial second harmonic with itself excites 4th, 6th, 8th, ... harmonics, 
but also the 0th harmonic.  In turn, these higher harmonics affect the non-linear response of $v_2$. For this reason,
numerical studies of the non-linear response to $v_2$ require sufficiently fine discretization in the momentum angle $\phi$ 
to follow numerically also the higher excited harmonics. The numerical results shown here were obtained for a $\phi$-range
discretized with 40 points, and their numerical stability was checked with finer discretizations. Our first main result is to observe
that the non-linearities are more important for large opacity, as the lines in Fig.~\ref{fig1} develop larger slopes and curvatures.
While the numerical results for $v_2$ and $v_4$  in Fig.~\ref{fig1}
do not involve a perturbative expansion in $\epsilon_n$ or $\hat\gamma$, symmetry arguments imply that they 
must agree for sufficiently small $\epsilon_2$ with the perturbative series
\begin{equation}
V_2 = 
w_{2,2} \epsilon_2 + w_{2,222}  \left( \epsilon_2  \epsilon_2^*\right) \epsilon_2 + O(\vert\epsilon_2\vert^5)\, ,
\qquad
V_4 = 
w_{4,22} \epsilon_2^2  + O(\vert\epsilon_2\vert^4)\, .
\label{eps2response}
\end{equation}
According to eq.~\eqref{eps2response},
the response coefficient $w_{2,2}(\hat\gamma)$ at a given opacity $\hat\gamma$ is
the intercept  $\lim_{\epsilon_2^2 \to 0}\textstyle\frac{v_2}{\epsilon_2}(\epsilon_2^2)$ of the corresponding
curve in Fig.~\ref{fig1} with the ordinate.  The non-linear response coefficicient $w_{2,222}(\hat\gamma) $ is
 the slope of the same curves in Fig.~\ref{fig1} at $\epsilon_2^2 = 0$. Similarly, one finds the non-linear response 
 $w_{4,22}(\hat\gamma) = \lim_{\epsilon_2^2 \to 0}\textstyle\frac{v_4}{\epsilon_2^2}(\epsilon_2^2)$.  
 For notational simplicity, we do not denote explicitly the phases of the eccentricities in the following 
 as these can be inferred easily from symmetry arguments. Fig.~\ref{fig2} shows
 the $\hat\gamma$-dependence of the linear response coefficient extracted from Fig.~\ref{fig1} in this way. 
 
\begin{figure}[t]
 \includegraphics[width=0.5 \textwidth]{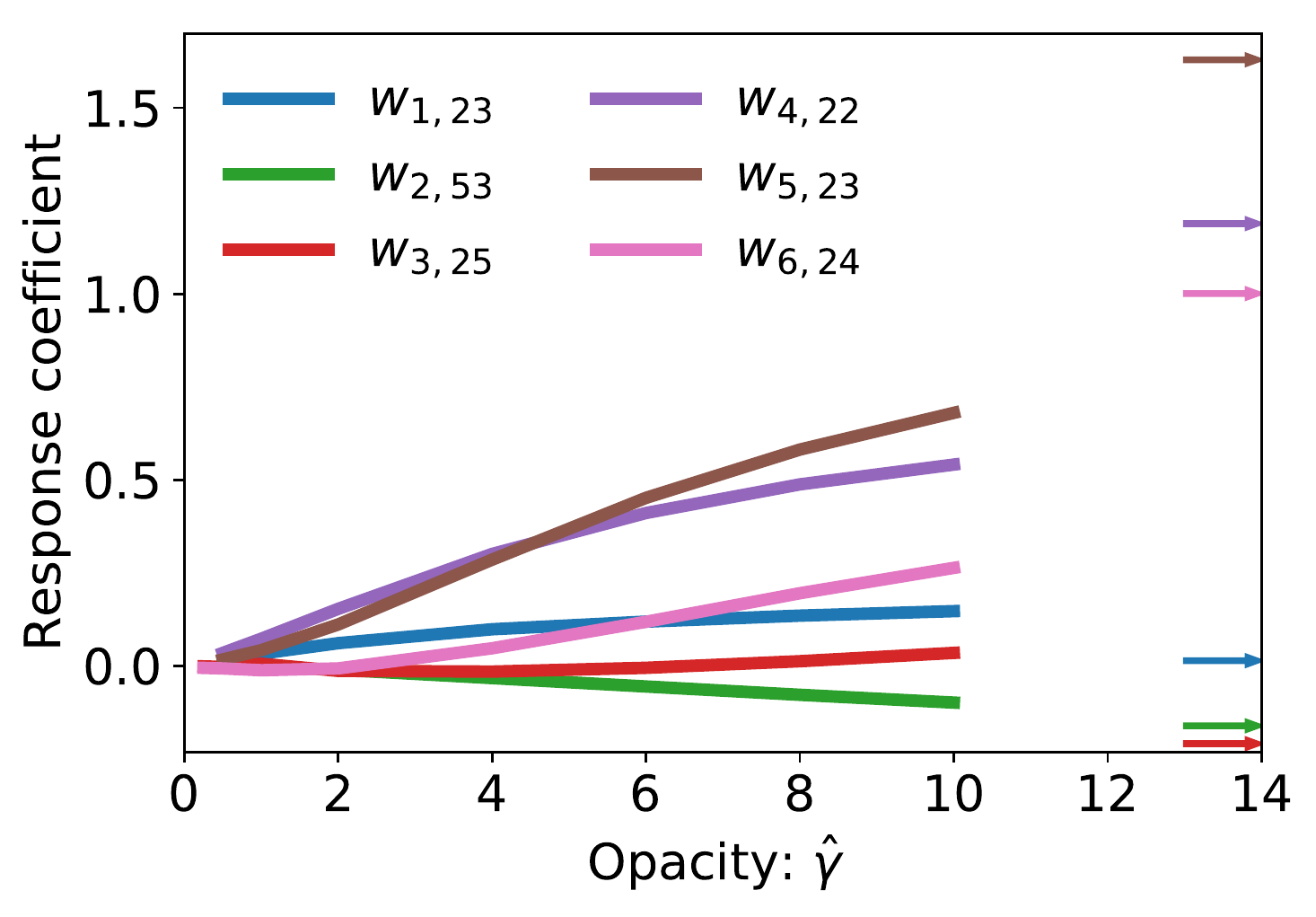}
 \includegraphics[width=0.5 \textwidth]{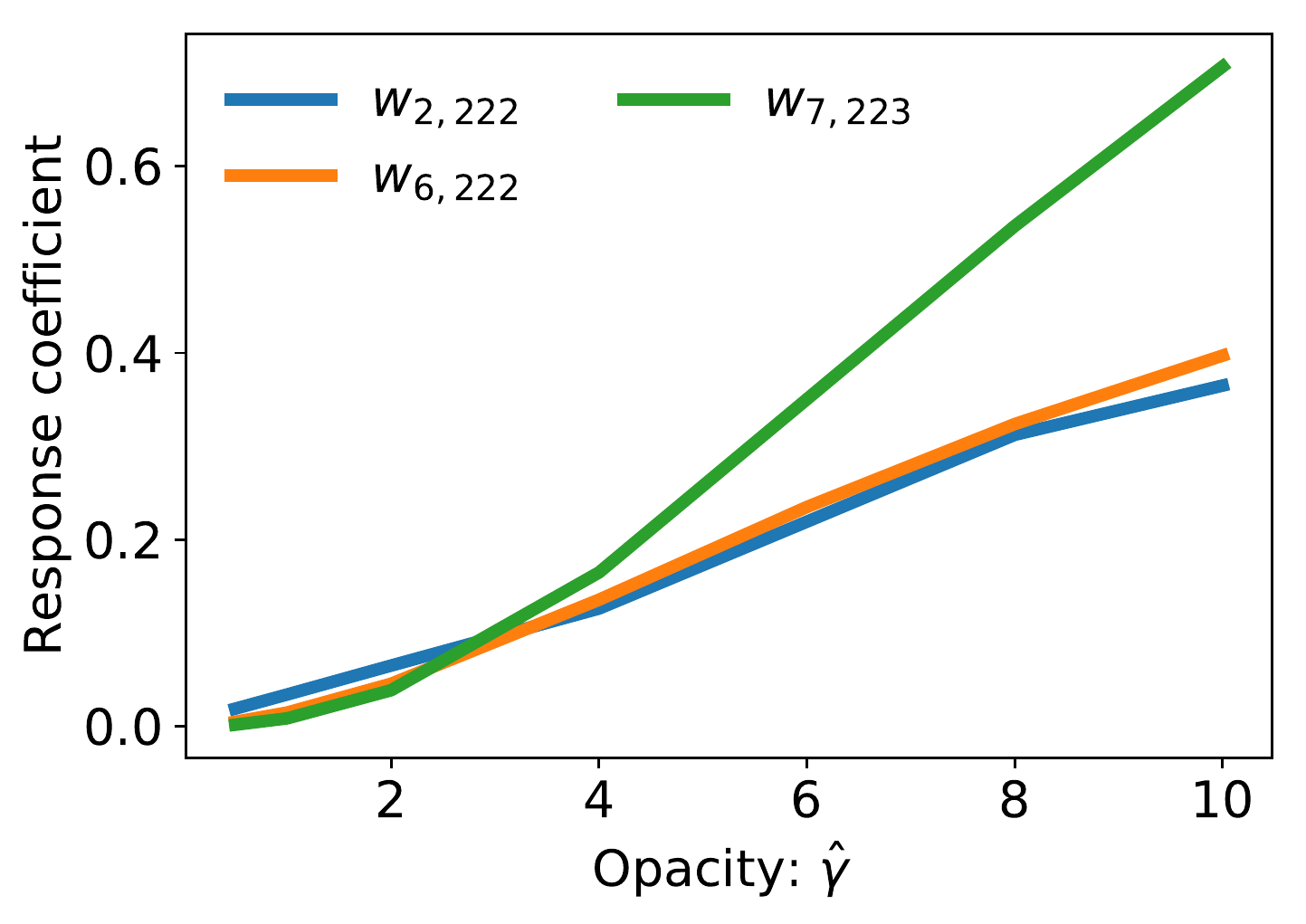}
  \caption{Left panel: Quadratic response coefficients calculated from the kinetic theory \eqref{eq1}, \eqref{eq2} (thick lines)
  and their ideal-fluid limits coresponding to $\hat\gamma \to \infty$ (arrows).  Right panel: Cubic response coefficients
  from the same kinetic theory.}
  \label{fig3}
\end{figure}

In close analogy, we determine other linear and non-linear response coefficients numerically by seeding the initial
conditions with suitable choices of eccentricities. To determine the linear response coefficients $w_{n,n}(\hat\gamma)$, $n\leq 5$, 
shown in Fig.~\ref{fig2}, we run simulations seeded with a single $n$-th harmonic for different values of 
$\epsilon_n$, and we extrapolate to $\lim_{\epsilon_n^2 \to 0}\textstyle\frac{v_n}{\epsilon_n}(\epsilon_n^2)$, see Fig.~\ref{fig2}.
For the non-linear response coefficients $w_{n, m_1 m_2}$ ($n= m_1+m_2$ or $= \vert m_1 - m_2\vert$),
displayed in Fig.~\ref{fig3}, we pick initial data with non-vanishing $\epsilon_{m_1}$, $\epsilon_{m_2}$ and all other
eccentricities vanishing. Extrapolating from simulations for different initial values of $\epsilon_{m_1}$, $\epsilon_{m_2}$, we determine
$w_{n,m_1 m_2} = \lim_{\epsilon_{m_1}, \epsilon_{m_2}\to 0} \textstyle\frac{v_n}{\epsilon_{m_1} \epsilon_{m_2}}$.

 We ask next how the linear and non-linear response coefficients in Figs.~\ref{fig2} and \ref{fig3} hydrodynamize, \emph{i.e.}, how they 
approach their fluid-dynamic limit with increasing opacity $\hat\gamma$. 
 To this end, we relate the opacity  that characterizes kinetic
 transport to quantities accessible in viscous fluid dynamics. The definition $\hat\gamma = \gamma\, R^{3/4}\,  \left(\varepsilon_0\, \tau_0 \right)^{1/4}$ assumes that the early-time evolution is given by free-streaming which is not the case for viscous fluid dynamics. 
 We therefore have to work with an equivalent definition that can be expressed in terms of quantities measured at a time at
 which the flow builds up and fluid dynamics may be operational. To this end, we write
\begin{equation}
\hat\gamma 
= \gamma\, R\, \left(\frac{\varepsilon_{R}}{f_{0\to R}(\hat\gamma)}\right)^{1/4}\, ,
\label{eq6}
\end{equation}
where, for the Gaussian background in the initial condition \eqref{eq2}, $\varepsilon_0$ and $\varepsilon_{R} $ denote 
central ($r=0$) energy densities at times $\tau_0$ and $R$, 
respectively. The function $f_{0\to R}(\hat\gamma) = \frac{\varepsilon_R R}{\varepsilon_0 \tau_0}$ is defined as the ratio of the energy per unit rapidity at time 
$\tau=R$ to the energy which the system would have if it were free-streaming \cite{Kurkela:2019kip}. We calculate $f_{0\to R}(\hat\gamma)$ from kinetic theory for $\hat\gamma \leq 10$, and we match for larger $\hat\gamma$ to the known 
asymptotic large-$\hat\gamma$ behavior $f_{0\to R} \sim \hat\gamma^{-4/9}$. 

With $f_{0\to R}(\hat\gamma)$ known, we relate viscous fluid-dynamic calculations to $\hat\gamma$ by specifying 
$\varepsilon_{R}$ and $\eta/s$ from fluid dynamics and solving eq.~\eqref{eq6} for $\hat\gamma$. 
In particular,
we use the kinetic relation between the interaction strength $\gamma$ and the shear viscosity  $\frac{\eta}{sT} = \textstyle\frac{1}{5 \gamma\varepsilon^{1/4}}$.
We then initialize at some initial time $\tau_0 < R$ the components of 
$T^{\mu\nu}$ from the same initial conditions~\eqref{eq2} as the kinetic theory and we evolve them with viscous fluid 
dynamics for varying $\textstyle\frac{ \eta}{s}$ and $\varepsilon_0$. This allows us to determine $\varepsilon_{R}$ and $\hat \gamma$, and to extract from the
transverse energy flow \eqref{eq4} at late times the energy-flow coefficients $v_n$. In general, these results depend on $\tau_0$.
That the $\tau_0 \to 0$-limit of $v_n(\hat\gamma)$ exists is a direct consequence of the fact that viscous
fluid dynamics, like kinetic theory, has a universal attractor solution at arbitrarily early times~\cite{Romatschke:2017vte}. While the attractor of kinetic theory keeps $\varepsilon \tau$ fixed leading to the scaling of $\hat\gamma = \gamma\, R^{3/4}\,  \left(\varepsilon_0\, \tau_0 \right)^{1/4}$ , the attractor of the viscous (Israel-Stewart) hydrodynamics considered here keeps $\varepsilon \tau^{\frac{4}{15} \left(\sqrt{5}-5\right)}$ constant~\cite{Kurkela:2019set}. Therefore taking the $\tau_0 \to 0$ limit 
while keeping $\hat\gamma $ as defined in eq.~\eqref{eq6} fixed corresponds to scaling initial energy densities by $\varepsilon(\tau_0) \propto \tau_0^{-\frac{4}{15} \left(\sqrt{5}-5\right)}$. While this non-standard procedure differs from the common phenomenological practise, it allows for a particularly clean comparison between kinetic theory and fluid dynamics by eliminating the unphysical model parameter $\tau_0$.  The difference between the kinetic theory and fluid-dynamic results obtained this way do not inform us on the validity or the breakdown of the current phenomenological practice. Instead it emphasizes the importance of the early-time attractor (which differs between kinetic theory and the fluid dynamics) for the physical observables measured in experiments and it informs us about the extent to which the entire signal $v_n$ is or is not build up by the degrees of freedom  encoded in viscous fluid dynamics.
For linear response coefficients, this comparison is shown in the right panel of Fig.~\ref{fig2}.



  Technically, we evolve the viscous fluid-dynamic equations as described in Ref.~\cite{Floerchinger:2013rya,Floerchinger:2013tya} 
  by splitting all fluid dynamic fields into an azimuthally symmetric background and an azimuthally anisotropic perturbation and solving
  for them to first order in initial eccentricites.  In the same way, we set up a control calculation for the much simpler ideal fluid-dynamic equations to obtain an independent determination of linear response coefficients
in the limit $\hat\gamma \to \infty$ (arrows in Fig.~\ref{fig2}). Results for $w_{2,2}$ and $w_{3,3}$ differ somewhat from those reported in \cite{Kurkela:2019kip} since the initial conditions are different.

 As expected from general reasoning, the viscous fluid-dynamic results for $\textstyle\frac{v_n(\hat\gamma)}{\epsilon_n} $ in 
 the limit $\tau_0 \to 0$ asymptote for $\hat\gamma \to \infty$ to the ideal fluid-dynamic results in the same 
  $\tau_0 \to 0$  limit, see Fig.~\ref{fig2}. 
  Remarkably, the hierarchy between the elliptic and triangular linear response coefficient gets inverted as a function of $\hat\gamma$: 
  kinetic theory at low $\hat\gamma$ shows $w_{2,2} > w_{3,3}$ while ideal fluid dynamics shows $w_{2,2} < w_{3,3}$.
  Viscous fluid dynamics accounts for this inversion qualitatively: for very small specific shear viscosity $\textstyle\frac{\eta}{s}$, 
  \emph{i.e.}, very large opacity $\hat\gamma$, it is consistent with ideal fluid dynamics, but the hierarchy changes as 
  a function of opacity, see right panel of Fig.~\ref{fig2}. Also the results from kinetic theory hint at such an inversion,
  as the slope of $w_{3,3}(\hat\gamma=10)$ is larger than the slope of $w_{2,2}(\hat\gamma=10)$.

As seen from Fig.~\ref{fig2}, viscous fluid dynamics reproduces the main qualitative trends of kinetic theory (hierarchy of response
coefficients) at $\hat\gamma \sim O(10)$, but significant quantitative differences persist. 
 On general grounds, we expect
that kinetic theory matches quantitatively to viscous fluid dynamics at sufficiently large $\hat\gamma$ when the 
fluid dynamic gradient expansion becomes quantitatively reliable. All data shown here are consistent with this expectation. 
It would clearly be interesting to extend the numerical calculations in kinetic theory to larger $\hat\gamma$ and to determine 
the $\hat\gamma$-scale at which a seamless matching to viscous fluid dynamics is found. However, with increasing 
$\hat\gamma$, the numerical evaluation becomes more expensive, and within the scope of the present letter, we were not 
able to push to higher $\hat\gamma$.

We have extended this analysis to a set of quadratic and cubic response coefficients, see Fig.~\ref{fig3}. 
To make some statements about their hydrodynamization 
we determine the quadratic response coefficients in the limit $\hat\gamma \to \infty$ by solving
ideal fluid dynamics to second order in eccentricities (arrows in Fig.~\ref{fig3}). Within the range $\hat\gamma < 10$,  
several quadratic response coefficients are seen to cross, and at $\hat\gamma = 10$, the hierarchy of the numerically 
large response coefficients ($w_{5,23} > w_{4,22} > w_{6,24}$) found in kinetic theory is consistent with that of ideal 
fluid dynamics. In the range  $\hat\gamma > 10$,  the numerically smaller response coefficients $w_{3,25}$ and $w_{2,53}$
need to cross. 
These  observations give further support to the conclusions reached from Fig.~\ref{fig2}. 

In a remarkable note \cite{Borghini:2018xum}, it was observed already 
that in the dilute limit of kinetic theory  far from equilibirum, linear and quadratic response coefficients 
grow linearly in the average number of rescatterings $\bar{N}_{\rm resc.}$ while cubic ones have a
quadratic dependence. In Ref.~\cite{Borghini:2018xum}, this scaling was established for 
elastic two-to-two collision kernels. The line of arguments of Ref.~\cite{Borghini:2018xum}  does not apply 
 to the collision kernel \eqref{eq1}. 
 However, a perturbative expansion of \eqref{eq1} in $\hat\gamma$ can be viewed as an expansion in the
average number of scattering centers~\cite{Kurkela:2018ygx}, and it is therefore natural to test whether our
results show this same scaling, too. For linear and quadratic coefficients, we know already from the perturbative
analysis in ~\cite{Kurkela:2018ygx} that they do. 
For cubic response coefficients, however, we observe small violations of 
the scaling. In the neighborhood of $\hat\gamma=0$, the cubic coefficients in the right panel of Fig.~\ref{fig3} 
show a small linear component, though the quadratic one can be dominant. 

\begin{figure}[t]
\hspace{-2cm}
   \includegraphics[width=1. \textwidth]{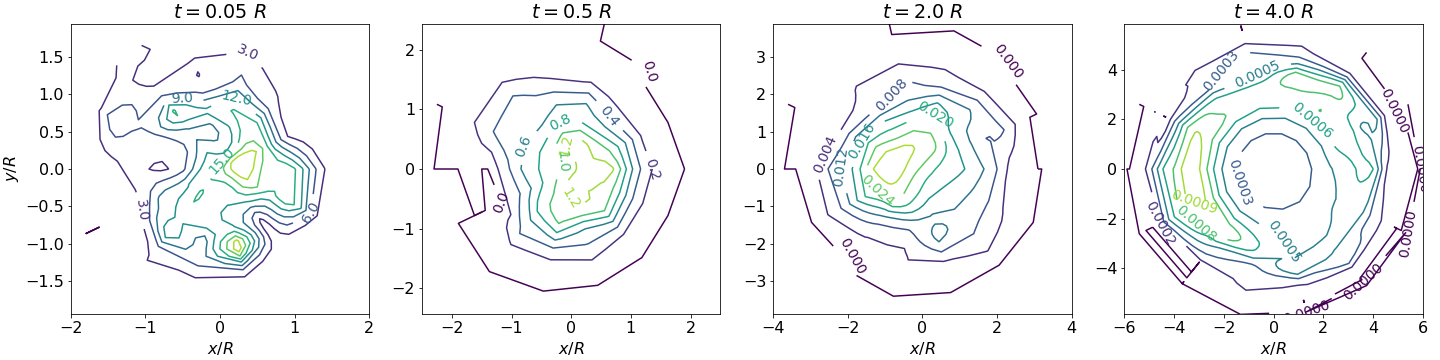}
 \includegraphics[width=0.45 \textwidth]{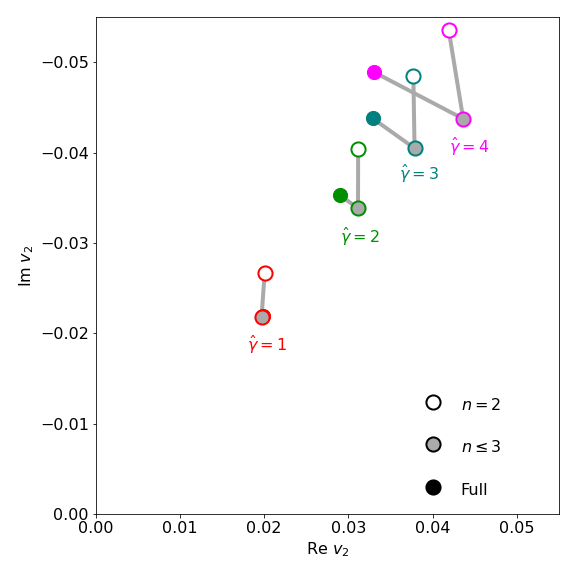}
 \includegraphics[width=0.45 \textwidth]{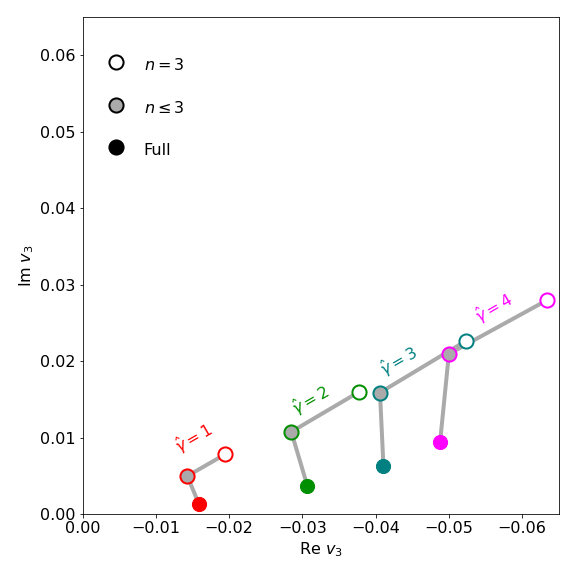}
  \caption{Upper panels: Energy density in the transverse plane initialized with the \Trento model at $\tau_0/R = 0.05$ and evolved for $\hat\gamma = 2$ 
  with the kinetic theory up to times $\tau/R = 0.05$, $0.5$, $2.0$ and $4.0$, respectively. Lower panel: The value of the elliptic and triangular flow coefficients evaluated for the same \Trento event and for different opacities $\hat\gamma$. Results for the \emph{full} event including harmonics $n\leq 7$ are compared to simplified events in which only specific harmonics are kept.
  (For $\hat\gamma = 1$, the circles for $n\leq3$ and $full$ event overlap in the left plot.)}
  \label{fig4}
\end{figure}

{\bf Evolving initial conditions with realistic event-by-event fluctuations in kinetic theory.} 
We now apply our newly developed numerical machinery to the first exploratory study of a realistic initial condition that
would be one single event in an event sample of an event-by-event analysis. The initial condition is a typical \Trento event~\cite{Moreland:2014oya}
in the $5-10 \%$ centrality class smoothened such that only initial $\epsilon_n$'s for $n \leq 7$ are kept. 
We have checked that the finesse of our
discretization allows for the stable propagation of such events. A typical time evolution is shown in the upper panel of Fig.~\ref{fig4} with $\hat\gamma = 2$.
It illustrates that the Boltzmann equation can be solved non-perturbatively for distribution functions representing realistic initial conditions. 

The radial profile of  the \Trento event studied here differs from \eqref{eq2} and this can affect the value of linear and non-linear
response coefficients. To quantify the difference, we compare the $w_{2,2}$ extracted for these two profiles and find the
following numbers $w_{2,2}^{\rm (TrENTo)} = 
\textstyle\frac{v_2}{\epsilon_2} \vert_{\rm Trento, n=2} = 0.156$, $0.239$, $0.288$ and $0.319$,
compared to $w_{2,2}(\hat\gamma) = 0.166$, $0.266$, $0.327$ and $0.372$ taken from Fig.~\ref{fig2} for
$\hat\gamma= 1,2,3, 4$. Technically, $ w_{2,2}^{\rm (TrENTo)} $ is not a linear response coefficient, since it was
extracted at finite eccentricity, but Fig.~\ref{fig1} informs us that the numerical contribution arising from finite eccentricity
is negligible for small opacity. We checked this for the \Trento profile as well (data not shown). We observe that 
the dependence on the radial profile in the linear response coefficients ranges from 5\% to 15\% in this $\hat\gamma$-range.
The analogous study of $w_{3,3}$ shows a  2\% to 10\% difference in the same $\hat\gamma$-range.
Therefore, the open circles in the lower panel of Fig.~\ref{fig4} are accounted for within 2\% - 15\% accuracy by 
the linear response coefficients calculated from the simplified profile \eqref{eq2}. 
The remaining difference between open circles and full results in Fig.~\ref{fig4} result from mode-mode couplings of different
harmonics. We see that while the linear response covers the ballpark of the results, non-linearities have to be included to 
go reliably beyond 20\%-30\% accuracy. The non-linearities generated by the lowest harmonics $n\leq 3$ account for 
half of all the non-linearities. 

This paper is motivated by the wealth of studies of hydrodynamization and thermalization in simplified settings. 
We have developed the necessary machinery for overcoming many of these simplications and 
to facilitate studies of hydrodynamization in complex realistic geometries, and to thus push the study of hydrodynamization 
from \emph{in vitro} to \emph{in vivo}. The ability to solve the Bolzmann equation for ultra-relativistic 
systems with realistic initial geometries and including all non-linear mode-mode couplings provides insight into 
how the characteristic features of fluid dynamics
 emerge gradually with increasing interaction strength. Away from the fluid dynamic limit, signals of collectivity are not simply 
 reduced uniformily in size, but their relative strength varies characteristically with opacity, the hierarchy of the dominant
 linear response coefficients is inverted and so is the hierarchy of several non-linear ones. This may provide novel possibilities
 for characterizing to what extent systems of different size do or do not hydrodynamize. 
   In the long run, we hope that the 
 technical advances documented here can be developed further to study the evolution of event samples, and to study 
 Boltzmann equations with other phenomenologically relevant complications. 

{\bf Acknowledgements.} One of us (SFT) has received funding from the European Research Council (ERC) under the European Unions Horizon 2020 research and innovation programme (grant agreement No 759257).

\end{document}